\documentstyle[12pt]{article}
\newcommand{\be}{\begin{equation}}
\newcommand{\ee}{\end{equation}}
\newcommand{\bea}{\begin{eqnarray}}
\newcommand{\eea}{\end{eqnarray}}
\newcommand{\un}{\underline}
\newcommand{\eps}{\varepsilon}
\newcommand{\zahlen}{{\rm Z \!\! Z}}
\newcommand{\half}{{\scriptstyle{{1\over 2}}}}
\newcommand{\twth}{{\scriptstyle{{1\over 12}}}}
\newcommand{\thalf}{{\scriptstyle{{3\over 2}}}}
\newcommand{\fhalf}{{\scriptstyle{{5\over 2}}}}
\newcommand{\fthird}{{\scriptstyle{{5\over 3}}}}
\newcommand{\eighth}{{\scriptstyle{{1\over 8}}}}
\newcommand{\sn}{{\scriptstyle{{1-}}}}
\newcommand{\real}{\relax{\rm I\kern-.18em R}}
\newcommand{\Tr}{{\rm Tr}}
\newcommand{\ad}{{\rm ad}}
\newcommand{\cF}{{\cal F}}
\newcommand{\cD}{{\cal D}}
\newcommand{\cP}{{\cal P}}
\newcommand{\cO}{{\cal O}}
\newcommand{\cE}{{\cal E}}

\newcommand{\basispl}{
   \put(-.5,-.5){\line(1,0){1}}
   \put(.5,-.5){\line(0,1){1}}
   \put(.5,.5){\line(-1,0){1}}
   \put(-.5,.5){\line(0,-1){1}}
                         }
\newcommand{\basisar}{
   \put(0,-.5){\vector(1,0){0}}
   \put(.5,0){\vector(0,1){0}}
   \put(0,.5){\vector(-1,0){0}}
   \put(-.5,0){\vector(0,-1){0}}
	              }
\newcommand{\revar}{
   \put(0,-.5){\vector(-1,0){0}}
   \put(.5,0){\vector(0,-1){0}}
   \put(0,.5){\vector(1,0){0}}
   \put(-.5,0){\vector(0,1){0}}
	              }
\newcommand{\plaq}{\setlength{\unitlength}{.5cm}\raisebox{-.2cm}{
   \begin{picture}(1.2,1.2)(-.6,-.6)
   \basispl\basisar
   \put(-.5,-.5){\circle*{.2}}
   \put(-.55,-.55){\makebox(0,0)[tr]{\footnotesize $x$}}
   \put(-.55,0){\makebox(0,0)[r]{\footnotesize $\nu$}}
   \put(0,-.55){\makebox(0,0)[t]{\footnotesize $\mu$}}
   \end{picture}}}
\newcommand{\twoplaq}{\setlength{\unitlength}{1cm}\raisebox{-.5cm}{
   \begin{picture}(1.2,1.2)(-.6,-.6)
   \basispl
   \put(-.5,-.5){\circle*{.1}}
   \put(-.5,.5){\circle*{.1}}
   \put(.5,-.5){\circle*{.1}}
   \put(.5,.5){\circle*{.1}}
   \put(0,-.5){\circle*{.1}}
   \put(0,.5){\circle*{.1}}
   \put(.5,0){\circle*{.1}}
   \put(-.5,0){\circle*{.1}}
   \put(-.25,-.5){\vector(1,0){0}}
   \put(.25,-.5){\vector(1,0){0}}
   \put(.5,-.25){\vector(0,1){0}}
   \put(.5,.25){\vector(0,1){0}}
   \put(-.25,.5){\vector(-1,0){0}}
   \put(.25,.5){\vector(-1,0){0}}
   \put(-.5,-.25){\vector(0,-1){0}}
   \put(-.5,.25){\vector(0,-1){0}}
   \put(-.55,-.55){\makebox(0,0)[tr]{\footnotesize $x$}}
   \put(-.55,0){\makebox(0,0)[r]{\footnotesize $\nu$}}
   \put(0,-.55){\makebox(0,0)[t]{\footnotesize $\mu$}}
   \end{picture}}}
\newcommand{\stapup}{\setlength{\unitlength}{.5cm}\raisebox{-.2cm}{
   \begin{picture}(1.2,1.2)(-.6,-.6)
   \put(.5,-.5){\line(0,1){1}}
   \put(.5,.5){\line(-1,0){1}}
   \put(-.5,.5){\line(0,-1){1}}
   \put(.5,0){\vector(0,-1){0}}
   \put(0,.5){\vector(1,0){0}}
   \put(-.5,0){\vector(0,1){0}}
   \put(-.5,-.5){\circle*{.2}}
   \put(-.55,-.55){\makebox(0,0)[tr]{\footnotesize $x$}}
   \put(-.55,0){\makebox(0,0)[r]{\footnotesize $\nu$}}
   \put(0,.55){\makebox(0,0)[b]{\footnotesize $\mu$}}
   \end{picture}}}
\newcommand{\stapdw}{\setlength{\unitlength}{.5cm}\raisebox{-.2cm}{
   \begin{picture}(1.2,1.2)(-.6,-.6)
   \put(.5,-.5){\line(0,1){1}}
   \put(.5,-.5){\line(-1,0){1}}
   \put(-.5,.5){\line(0,-1){1}}
   \put(.5,0){\vector(0,1){0}}
   \put(0,-.5){\vector(1,0){0}}
   \put(-.5,0){\vector(0,-1){0}}
   \put(-.5,.5){\circle*{.2}}
   \put(-.55,.75){\makebox(0,0)[tr]{\footnotesize $x$}}
   \put(-.55,0){\makebox(0,0)[r]{\footnotesize $\nu$}}
   \put(0,-.55){\makebox(0,0)[t]{\footnotesize $\mu$}}
   \end{picture}}}
\newcommand{\clover}{\setlength{\unitlength}{.5cm}\raisebox{-.5cm}{
   \begin{picture}(2.4,2.4)(-1.2,-1.2)
   \multiput(-1.2,-1.2)(1.2,1.2){2}{\begin{picture}(1.2,1.2)(-.6,-.6)
   \basispl\basisar\end{picture}}
   \multiput(-1.2,0)(1.2,-1.2){2}{\begin{picture}(1.2,1.2)(-.6,-.6)
   \basispl\revar\end{picture}}
   \put(-.1,-.1){\circle*{.2}}
   \put(-.1,.1){\circle*{.2}}
   \put(.1,-.1){\circle*{.2}}
   \put(.1,.1){\circle*{.2}}
   \end{picture}}}
\newcommand{\twooneplaq}{\setlength{\unitlength}{.5cm}
   \raisebox{-.2cm}{
   \begin{picture}(2.2,1.2)(-1.1,-.6)
   \put(-1,-.5){\line(1,0){2}}
   \put(-1,.5){\line(1,0){2}}
   \put(-1,-.5){\line(0,1){1}}
   \put(1,-.5){\line(0,1){1}}
   \multiput(-1,-.5)(1,0){3}{\circle*{.2}}
   \multiput(-1,.5)(1,0){3}{\circle*{.2}}
   \end{picture}}}
\newcommand{\plaqa}{\setlength{\unitlength}{.5cm}\raisebox{-.2cm}{
   \begin{picture}(1.2,1.2)(-.6,-.6)
   \basispl
   \put(-.5,-.5){\circle*{.2}}
   \put(-.5,.5){\circle*{.2}}
   \put(.5,-.5){\circle*{.2}}
   \put(.5,.5){\circle*{.2}}
   \end{picture}}}
\newcommand{\hookplaq}{\setlength{\unitlength}{.5cm}
   \raisebox{-.3268cm}{
   \begin{picture}(1.7071,1.7071)(-.7071,-.7071)
   \put(0,0){\line(0,1){1}}
   \put(0,1){\line(1,0){1}}
   \put(1,1){\line(0,-1){1}}
   \put(-.7071,-.7071){\line(1,0){1}}
   \put(0,0){\line(-1,-1){.7071}}
   \put(1,0){\line(-1,-1){.7071}}
   \multiput(0,0)(1,0){2}{\circle*{.2}}
   \multiput(0,1)(1,0){2}{\circle*{.2}}
   \multiput(-.7071,-.7071)(1,0){2}{\circle*{.2}}
   \multiput(0,0)(.25,0){4}{\circle*{.03}}
   \end{picture}}}
\newcommand{\cornplaq}{\setlength{\unitlength}{.5cm}
   \raisebox{-.3268cm}{
   \begin{picture}(1.7071,1.7071)(-.7071,-.7071)
   \put(-.7071,-.7071){\line(0,1){1}}
   \put(0,1){\line(1,0){1}}
   \put(1,1){\line(0,-1){1}}
   \put(-.7071,-.7071){\line(1,0){1}}
   \put(0,1){\line(-1,-1){.7071}}
   \put(1,0){\line(-1,-1){.7071}}
   \put(-.7071,-.7071){\circle*{.1}}
   \put(-.7071,.2929){\circle*{.2}}
   \multiput(0,0)(1,0){2}{\circle*{.2}}
   \multiput(0,1)(1,0){2}{\circle*{.2}}
   \multiput(-.7071,-.7071)(1,0){2}{\circle*{.2}}
   \multiput(0,0)(.25,0){4}{\circle*{.03}}
   \multiput(0,0)(0,.25){4}{\circle*{.03}}
   \multiput(0,0)(-.1768,-.1768){4}{\circle*{.03}}
   \end{picture}}}

\def\phm{\hphantom{-}}
\def\mystrut{{\vrule height 15pt depth 4pt width 0pt}}

\begin{document}
\vskip-1cm
\hfill INLO-PUB-11/93, FTUAM-93/31
\vskip5mm
\begin{center}
{\LARGE{\bf{\underline{Instantons from over-improved cooling}}}}\\
\vspace*{1cm}{\large
Margarita Garc\'{\i}a P\'erez$^{(a)}$, Antonio Gonz\'alez-Arroyo$^{(b)}$,
Jeroen Snippe$^{(a)}$\\ and\\ Pierre van Baal$^{(a)}$\\}
\vspace*{1cm}
(a) Instituut-Lorentz for Theoretical Physics,\\
University of Leiden, PO Box 9506,\\
NL-2300 RA Leiden, The Netherlands.\\ 
\vspace*{5mm}
(b) Departamento de F\'{\i}sica Te\'{o}rica C-XI,\\
Universidad Aut\'{o}noma de Madrid,\\
28049 Madrid, Spain.
\end{center}
\vspace*{5mm}{\narrower\narrower{\noindent
\underline{Abstract:} Lattice artefacts are used, through
modified lattice actions, as a tool to find the largest
instantons in a toroidal geometry $[0,L]^3\times[0,T]$ for $T
\rightarrow \infty$. It is conjectured that the largest instanton
is associated with tunnelling through a sphaleron. Existence of
instantons with at least 8 parameters can be proven with the help
of twisted boundary conditions in the time direction. Numerical
results for $SU(2)$ gauge theory obtained by cooling are
presented to demonstrate the viability of the method.}\par}

\section{Introduction}

Since the time of the discovery of instantons~\cite{bel1} in non-abelian
gauge theories, as vacuum to vacuum quantum mechanical tunnelling
events~\cite{tho2}, their role in strongly interacting theories has been
controversial, both in the continuum~\cite{cal3} and in the lattice
formulation~\cite{kro4}. For the continuum this has been mainly due to
applying semiclassical techniques, which cannot be justified at
strong coupling. In the lattice formulation the main problems
were the instantons localised at the scale of the lattice cut-off
for which topological charge cannot be defined unambiguously~\cite{lus5},
and which have actions considerably lower than the continuum
action of $8 \pi^2$. Strictly speaking, there are no locally
stable solutions on a lattice using the standard Wilson action~\cite{wil29},
because this lattice action decreases when the instanton
becomes more localized~\cite{tep6}, as we will demonstrate also from
analytic considerations. On a trial and error basis, different
(improved) lattice actions were considered, some of them indeed
giving rise to stable lattice solutions~\cite{iwa7}. This paper will
provide the proper framework to understand the stability.

It is not too difficult to understand the reason of the
instability. At finite lattice sizes the lattice action deviates
from the continuum and this deviation is larger for stronger
fields. For the Wilson action, as we will show, the lattice
artefacts make the action decrease as compared to the continuum.
In the continuum, instantons have a scale (or size) parameter
$\rho$, on which the action does not depend. But the smaller
$\rho$ becomes, the larger the fields get, which makes the
lattice action \underline{decrease}. On dimensional grounds one
easily argues that (generically) $S_{latt}(a,\rho)=8\pi^2(1+(a/\rho)^2
d_2+\cO (a/\rho)^4)$ for $\rho\gg a$, which will be demonstrated in more
detail further on. For the Wilson action~\cite{wil29} $d_2<0$,
explaining the instability. Hence, one simply modifies the
action, such that $d_2>0$, in order to get stable solutions for
the maximal value of $\rho$ allowed by the volume $[0,L]^3$,
which is kept finite. As we are interested in the classical
solutions to the equations of motion, the modified action need not
be of the type of an improved action~\cite{sym8}, for which typically one
wants to achieve $d_2=0$, as in that case (as we will show) the
$(a/\rho)^4$ term might still destabilize the solution.

We deliberately want to keep $d_2 > 0$, which we will hence call
over-improvement. The reason is, that our motivation for embarking
on this project was to find the instantons with the largest scale
$\rho$. This presumably will correspond to tunnelling over the
lowest energy barrier, separating two classical vacua. The
configuration that corresponds to the lowest barrier height is
then conjectured to be a sphaleron (which exists due to the fact
that we keep the volume finite). A sphaleron~\cite{kli9} is by definition
a saddle point of the energy functional with precisely one
unstable direction, which corresponds to the direction of
tunnelling. In this way we use the instantons to map out the
part of the energy functional relevant for the dynamical region
where a semiclassical analysis of tunnelling amplitudes will
break down. We refer to a pilot study~\cite{baa10} on $S^3\times\real$ for
readers interested in this issue, and for an explanation of the
relevance of the geometry $T^3\times\real$ which is studied in
this paper. This geometry allows us to find the instantons using
the lattice approximation. For simplicity we restrict ourselves
to $SU(2)$ pure gauge theories.

\section{On the existence of continuum solutions}

The geometry $T^3\times\real$, in particular in a lattice
formulation, can be seen as a limiting case of an asymmetric four
torus $[0,L]^3\times[0,T]$. The only known solutions have
constant curvature~\cite{tho11} and hence cannot correspond to vacuum to
vacuum tunnelling, furthermore their topological charge is at
least 2.  Actually, it can be proven rigorously~\cite{bra12}, that for
$T$ finite, no regular charge 1 self-dual solutions can exist on
a four-torus (we will illustrate this with our numerical
results). As soon as we allow for twisted boundary conditions~\cite{tho13},
existence of minimal non-trivial topological charge
instanton solutions can be proven. One distinguishes two cases,
depending on the properties of the twist tensor $n_{\mu\nu}\in\zahlen_2$.

When $\eighth\epsilon_{\mu\nu\lambda\sigma}n_{\mu\nu}
n_{\lambda\sigma}=1\,\mbox{mod}\,2$, the topological charge is half-integer.
The minimal action allowed by the topological bound is therefore $4\pi^2$,
corresponding  to topological charge 1/2. 
As twist is also well defined on the lattice~\cite{gro15},
and in the above situation (called non-orthogonal twist) does not
allow for zero-action configurations, these instantons cannot
``fall through the lattice''. Indeed, the index theorem predicts
in this case 4 parameters
(8 $\times$ topological charge), which have to correspond to the
position parameters. The charge 1/2 instanton hence has fixed
size and cannot shrink due to lattice artefacts. Impressively
accurate results~\cite{gar17} were obtained for this case using the well
known cooling method~\cite{tep6,ber21} to find a solution of the (lattice)
equations of motion, whose smoothness and scaling with the
lattice volume leaves no room to doubt it provides an accurate
approximation to the continuum solution with action $4\pi^2$. In the continuum,
existence of smooth non-trivial (but not necessarily self-dual) solutions
was proven by Sedlacek~\cite{sed14}, whereas theorem 3.2.1. of 
ref.\cite{bra18a} states that the moduli space of self-dual solutions 
is isomorphic with a four-torus.

When $\eighth\epsilon_{\mu\nu\lambda\sigma}n_{\mu\nu}
n_{\lambda\sigma}=0\,\mbox{mod}\,2$, also called an orthogonal
twist, there are ``twist eating''~\cite{gro15} configurations, i.e.
configurations that have zero action \underline{and} are
compatible with twisted boundary conditions (see also~\cite{amb31}). For
$SU(2)$, it is not too difficult to show that as long as
$n_{\mu\nu} \neq 0\,\mbox{mod}\,2$ for some $\mu$ and $\nu$, this twist
eating configuration is unique~\cite{baa16}, up to a global gauge
transformation if a twist is introduced as in ref.~\cite{gro15,gar17}
and multiplication with elements of the center of the gauge group. With
twisted boundary conditions as originally defined by 't Hooft~\cite{tho13},
such a global gauge transformation would even change the
boundary conditions, and as $SO(3)$ bundles the twist eating configuration
is unique (For SU(N) it can be proven~\cite{baa37} that out of the $N^4$ 
center elements that can multiply the twist, only $N^2$ give rise to gauge 
inequivalent configurations). Under this condition it can be
shown~\cite{bra18a} that there are instanton solutions with 8 parameters
(its moduli space, when dividing out the trivial translation
parameters, is even related to a $K3$ surface~\cite{bra18a,bra18b}) using 
Taubes'~\cite{tau19} technique of glueing a localized instanton (with scale,
position and global gauge parameters) to the ``twist eating'' flat connection 
(i.e. zero action configuration). As the latter is \underline{not} invariant 
under global gauge transformations, the global gauge parameters of the
localized instantons are genuine parameters of the moduli space
(see also~\cite{don30}).

The reason twisted boundary conditions are useful, is that at
finite $T$ there are no exact instantons on $T^4$ with periodic
boundary conditions, but there are exact solutions for any
non-trivial twist in the time direction. As $T\rightarrow\infty$
these solutions are also solutions on $T^3\times\real$. This comes
about as follows. Since at $T\rightarrow\infty$ the action can
only stay finite if for $|t|\rightarrow\infty$ the energy
density goes to zero, we deduce from a vanishing magnetic energy
that up to a gauge  
\be
A_i(\vec{x},t\rightarrow\pm\infty)=C_i^{\pm}\sigma_3/2L\quad,\label{eq:toron}
\ee
where $C_i^{\pm}\in[0,4\pi]~(A_0=0)$ parametrizes the vacuum or 
toron valley~\cite{baa20}, whose gauge invariant observables are best described 
by the Polyakov line expectation values
\be
P_i\equiv\half\Tr\left({\rm P}\exp(i\int_0^L
A_i(\vec{x},t)dx_i\right)=\cos\left(C_i/2\right)\quad,\label{eq:pol}
\ee
(for the proper definiton in the presence of twist, see ref.\cite{gar17}.)
In the vacuum valley $P_i$ is space independent and the vanishing
of the electric energy at $t\rightarrow\pm\infty$ also
requires $P_i\quad(\mbox{or}\quad C^{\pm}_i)$ to be asymptotically time
independent. Instanton solutions on $T^3\times\real$ are hence characterized 
by the boundary conditions $C^{\pm}_i$ at $t\rightarrow\pm\infty$.
It is these general instantons that are physically relevant. 
It is not clear if solutions exist with arbitrary boundary values. 
Approaching $T\rightarrow\infty$, by using periodic boundary conditions 
(which would impose $C^+_i=C^-_i\,\mbox{mod}\,4\pi$ up to a periodic gauge
transformation) does not allow us to prove existence. As long as
$T$ is finite there are no solutions~\cite{bra12} and the proof of 
non-existence breaks down as $T\rightarrow\infty$. On the other
hand, with twist in the time direction, $n_{0i}=1$, even at $T$
finite there is in the continuum an 8 parameter set of exact
instanton solutions, which at $T\rightarrow\infty$ will
correspond to $C^+_i=(2\pi-C^-_i)\,\mbox{mod}\,4\pi$ (again
up to a periodic gauge transformation). For localized instantons,
asymptotically the field has to coincide with the unique flat
connection, which fixes the possible values of $C^\pm_i$ to
$\pi$, but at the other extreme, as the instanton in the
spatial direction extends up to the ``boundary'' of the torus, the
regions $t\rightarrow+\infty$ and $t\rightarrow-\infty$
no longer are connected, which will relax the fact that
$P_i=0$ $(C^{\pm}_i=\pi)$. Although we have no proof, it is
reasonable to assume that the 8 parameters for the instantons
close to the maximal size are described by $\rho$, the 4
position parameters and the 3 vacuum valley parameters $C^+_i$
or $C^-_i)$. Note that for $P_i\rightarrow 0$ as
$t\rightarrow\pm\infty$, the solution is both compatible
with twisted \underline{and} periodic boundary conditions
\underline{at infinite $T$}. In any case we have now learned that
on $T^3\times\real$ (i.e. with free boundary conditions at $t
\rightarrow\pm\infty$) there are, at least 8 and at most 11
continuous parameters that describe the instanton solutions for
vacuum to vacuum tunnelling.

\section{The lattice actions and cooling}

Let us start with discussing the standard Wilson action~\cite{wil29}
\be
S=\sum_{x,\mu,\nu}\Tr\left(1-\plaq\right)=\sum_{x,\mu,\nu} 
\Tr(1-U_\mu(x)U_\nu(x+\hat{\mu})U_\mu^{\dagger}(x+\hat{\nu})
U_\nu^{\dagger}(x))\quad,\label{eq:wac}
\ee
where $U_\mu(x)$ are $SU(2)$ group elements on the link that runs from $x$ to
$x+\hat{\mu}$, the latter being the unit vector in the $\mu$ direction.
To derive the equations of motion, we observe that $S$ depends on $U_\mu(x)$
through the expression:
\be
S(U_\mu(x))=\Tr(1-U_\mu(x)\tilde{U}_\mu^{\dagger}(x))+
\Tr(1-U_\mu^\dagger(x)\tilde{U}_\mu(x))\quad,\label{eq:wacu}
\ee\vskip-1mm\noindent
where
\be
\tilde{U}_\mu(x)=\sum_{\nu\neq\mu}\left(\stapup+\stapdw\right)=
\sum_{\nu\neq\mu}(U_\nu(x)U_\mu(x+\hat{\nu})U_\nu^{\dagger}(x+\hat{\mu})  
+U_\nu^{\dagger}(x-\hat{\nu})U_\mu(x-\hat{\nu})U_\nu(x+\hat{\mu}-\hat{\nu}))
,\label{eq:utilde}
\ee
which is independent of $U_\mu(x)$. Hence, $S(e^{X}U_\mu(x))-S(U_\mu(x))=
\cO (X^2)$ for any Lie algebra element $X$, implies
\be
\Tr[\sigma_i(U_\mu(x)\tilde{U}_\mu^{\dagger}(x)-\tilde{U}_\mu(x)
U_\mu^{\dagger}(x))]=0\quad,\label{eq:coolcond}
\ee
where $\sigma_i$ are the Pauli matrices. This is easily seen to imply that
$U_\mu(x)\tilde{U}_\mu^{\dagger}(x)$ is a multiple of the identity, and as 
$\tilde{U}_\mu$ is the sum of SU(2) matrices, it can be written as 
$\tilde{U}_\mu=a_0+i\vec{a}\cdot\vec{\sigma}$, with $a_\mu\in\real^4$. 
If we define $\|\tilde{U}_\mu\|=\sqrt{a_\mu^2}$, eq.(\ref{eq:coolcond}) is 
seen to imply
\be
U_\mu(x)=\pm\tilde{U}_\mu(x)/\|\tilde{U}_\mu(x)\|\quad.\label{eq:update}
\ee
As we are only interested in stable solutions (i.e. local minima of the 
action), the plus sign in eq.(\ref{eq:update}) is the relevant one. 
The process of iteratively finding the solution to the equations of motion 
is called cooling~\cite{ber21}, as in all cases it is devised such that
the action is lowered after each iteration. The easiest is to simply choose 
$U_\mu'(x)=\tilde{U}_\mu(x)/\|\tilde{U}_\mu(x)\|$
since the fixed point of this iteration is clearly a solution to the 
equations of motion. An optimal way to sweep through the lattice is to 
divide for each $\mu$ the links $U_\mu(x)$ in two mutually exclusive 
checkerboard patterns $\Pi_\mu^i$ such that all links on a particular 
pattern $\Pi_\mu^i$ (i.e. for fixed $i$ and $\mu$) can be changed 
simultaneously, which is a well known trick to vectorize this procedure. 
At the cost of roughly a factor two in memory-use, vectorization is also 
achieved for the modified action we have considered so far for our numerical 
simulations:
\be
S(\eps)=\frac{4-\eps}{3}\sum_{x,\mu,\nu}\Tr\left(1-\plaq\right)
+\frac{\eps-1}{48}
\sum_{x,\mu,\nu}\Tr \left(1-\twoplaq\right)\quad.\label{eq:epsac1}
\ee
The meaning of the parameter $\eps$ will become clear in the next section. 
For ease of our numerical studies we have not considered modified single 
plaquette actions (see also the next section for a discussion on the 
adjoint and Manton actions).

%
\section{Lattice artefacts}

To calculate the effect of the discretization on the solutions of the equations 
of motion we first take a smooth continuum configuration (not necessarily a 
solution) $A_\mu(x)$. For definiteness we put $L=1$, and $N_s$ the number of 
lattice points in the spatial direction such that $a=1/N_s$. We put this 
configuration on the lattice by defining:
\be
U_\mu(x)={\rm Pexp}(i\int_0^a A_\mu(x+s\hat{\mu})ds)\quad.\label{eq:link}
\ee
The value of the plaquette thus corresponds to parallel transport around a 
square and can easily be proven to be given by~\cite{baa22} 
($D_\mu=\partial_\mu+A_\mu(x)$ 
the covariant derivative in the fundamental representation)
\be
\Tr\left(\plaq\right)=\Tr(e^{aD_\mu(x)}e^{aD_\nu(x)}e^{-aD_\mu(x)}
e^{-aD_\nu(x)})\quad.\label{eq:plaquette}
\ee
The proof simply amounts to observing that if $A_\mu(x)=A_\mu$, i.e. $A_\mu$ 
is space-time independent, then $\Tr\left(\plaq\right)=\Tr(e^{aA_\mu}e^{aA_\nu}
e^{-aA_\mu}e^{-aA_\nu})$ and eq.(\ref{eq:plaquette}) is the only way to 
make this formula gauge invariant under arbitrary (i.e. $x$-dependent)  
gauge transformations. Using the Campbell-Baker-Hausdorff formula,
eq.(\ref{eq:plaquette}) can be expressed in terms of products of 
covariant derivatives $\cD_\mu$ (in the adjoint representation) acting on the 
curvature $F_{\mu\nu}\equiv[D_\mu,D_\nu]=\partial A_\mu-\partial A_\nu+
[A_\mu,A_\nu]$, e.g. $\cD_\mu F_{\mu\nu}=[D_\mu,[D_\mu,D_\nu]]$. 
As the action involves a sum over all $x$, $\mu$ and $\nu$, things can be 
considerably simplified by computing, what we will call, the clover average
\bea
\left\langle\Tr\left(\plaq\right)\right\rangle_{\rm clover}
&=&\frac{1}{4}\Tr\left(\clover\right)
=\frac{1}{4}\Tr[e^{-aD_\mu}e^{-aD_\nu}e^{aD_\mu}e^{aD_\nu}
+e^{-aD_\mu}e^{aD_\nu}e^{aD_\mu}e^{-aD_\nu} \nonumber\\
& &\hskip3.5cm+e^{aD_\mu}e^{-aD_\nu}e^{-aD_\mu}e^{aD_\nu}
+e^{aD_\mu}e^{aD_\nu}e^{-aD_\mu}e^{-aD_\nu}]\nonumber\\
&=&\Tr[1+\frac{a^4}{2}F_{\mu\nu}^2(x)-\frac{a^6}{24}\left((\cD_\mu 
F_{\mu\nu}(x))^2+(\cD_\nu F_{\mu\nu}(x))^2\right)+\frac{a^8}{24}\{F_{\mu\nu}^4
(x)\nonumber\\& &+\frac{1}{30}\left((\cD_\mu^2 F_{\mu\nu}(x))^2+
(\cD_\nu^2 F_{\mu\nu}(x))^2\right)+\frac{1}{3}\cD_\mu^2 F_{\mu\nu}(x)
\cD_\nu^2 F_{\mu\nu}(x)\nonumber\\& &-\frac{1}{4}(\cD_\mu \cD_\nu 
F_{\mu\nu}(x))^2\}]+\cO (a^{10})+\mbox{total derivative terms}\quad,
\label{eq:clover}
\eea
for which the multiple Campbell-Baker-Hausdorff expansion of 
eq.(\ref{eq:plaquette}) is required to $\cO(a^6)$, obtained with the 
aid of the symbolic manipulation program FORM~\cite{ver23}. 
The clover average allows one to ignore many terms (all 
those odd in any of the indices) in evaluating the trace of the exponent.

Equation (\ref{eq:clover}) was also derived using the non-abelian Stokes 
formula~\cite{are24} ($s_0\equiv1$)
\bea
&&\hskip-5mm U_{\mu\nu}(x)\equiv U_\mu(x)U_\nu(x+\hat{\mu})
U_\mu^{\dagger}(x+\hat{\nu})U_\nu^{\dagger}(x)={\rm Pexp}( a^2\int_0^1
ds\int_0^1 dt~\cF_{\mu\nu}(x+a s\hat{\mu}+a t\hat{\nu}))\nonumber \\
&&\hskip-5mm\equiv1+\sum_{n=1}^\infty\prod_{i=1}^n\int_0^{s_{i-1}}ds_i\int_0^1
dt_i~a^2\cF_{\mu\nu}(x+a s_1\hat{\mu}+a t_1\hat{\nu})\ldots
a^2 \cF_{\mu\nu}(x+a s_n\hat{\mu}+a t_n\hat{\nu})\ ,\label{eq:stokes}
\eea
where $\cF_{\mu\nu}(y)$ equals $F_{\mu\nu}(y)$ up to the backtracking 
loop that connects $y$ to $x$, or:
\bea
& &V(s,t)={\rm Pexp}(a\int_0^sA_\mu(x+a\tilde{s}\hat{\mu})d\tilde{s})
{\rm Pexp}(a\int_0^tA_\mu (x+a s\hat{\mu}+a\tilde{t}\hat{\nu})d\tilde{t})
\quad,\nonumber \\
& &\hskip2cm\cF_{\mu \nu}(x+a s\hat{\mu}+a t\hat{\nu})=V(s,t) 
F_{\mu\nu}(x+a s\hat{\mu}+a t\hat{\nu})V^{\dagger}(s,t)\quad.
\label{eq:backtrack} 
\eea
To obtain the result of eq.(\ref{eq:clover}) one now expands 
$\cF_{\mu \nu}(x +a s\hat{\mu}+a t\hat{\nu})$ around the point 
$x$, making use of the identity:
\be
\partial_\mu^n\partial_\nu^m\cF_{\mu\nu}(x)=\cD_\mu^n\cD_\nu^m 
F_{\mu\nu}(x)\quad.\label{eq:der}
\ee
Note that the ordering of the covariant derivatives in the r.h.s. of 
eq.(\ref{eq:der}) is essential. Also crucial is that the path ordering
$U(s,t)\equiv{\rm Pexp}(\int_s^t A(u)du)$ (where $A(t)=\hat{e}_\mu A_\mu 
(x+t\hat{e})$ for some unit vector $\hat{e}$) is compatible with the 
covariant derivative, i.e. $\hat{e}_\mu D_\mu (x+s\hat{e})U(s,t)=0=
\hat{e}_\mu D_\mu (x+t\hat{e})U^{\dagger}(s,t)$ (in this respect we have
corrected the formula in ref.\cite{are24}). Inserting the Taylor expansion of
$\cF_{\mu \nu}(x+a s\hat{\mu}+a t\hat{\nu})$ with respect to $(s,t)$ in 
eq.(\ref{eq:stokes}), gives the result of eq.(\ref{eq:clover}). A very useful 
check is that the symmetry implied by $U_{\mu\nu}(x)=U_{\nu\mu}^{\dagger}(x)$,
not explicit at intermediate steps of the calculation, is respected by the 
final result.

Using eq.(\ref{eq:link},\ref{eq:clover}), one finds to $\cO (a^{10})$ for the 
modified action $S(\eps)$
\bea
& &\hskip-1cm S(\eps)=\sum_{x,\mu,\nu}\Tr[-\frac{a^4}{2}F_{\mu\nu}^2+  
\frac{\eps a^6}{24}\left((\cD_\mu F_{\mu\nu}(x))^2+(\cD_\nu F_{\mu\nu}(x))^2
\right)-\frac{(15\eps-12)a^8}{72}\{F_{\mu\nu}^4(x)\nonumber\\&&\hskip-5mm
+\frac{1}{30}\left(\cD_\mu^2 F_{\mu\nu}(x))^2+(\cD_\nu^2 F_{\mu\nu}(x))^2\right)
+\frac{1}{3}\cD_\mu^2 F_{\mu\nu}(x)\cD_\nu^2F_{\mu\nu}(x)
-\frac{1}{4}(\cD_\mu \cD_\nu F_{\mu\nu}(x))^2\}].\label{eq:epsac2}
\eea
Obviously, $S(\eps=1)$ corresponds to the Wilson action, and the sign
of the leading lattice artefacts are simply reversed by changing the sign 
of $\eps$. Most of the numerical results were obtained for
$\eps=-1$, but $\eps$ is useful in the initial cooling
from a random configuration. By keeping $\eps>0$ as long as 
$S>8\pi^2$, and only switching to $\eps=-1$ when $S\sim 8\pi^2$, we can 
avoid the solution to get stuck at higher topological charges. 
Once we set $\eps=-1$, we have yet to see an instanton fall
 through the lattice. We will come back to these issues when 
discussing the numerical results. 
Also note that, as ${\rm Tr_\ad}(U)=|\Tr(U)|^2-1$, one finds that 
the Wilson action in  the adjoint representation $(S_{\rm ad})$ satisfies 
$S_{\rm ad}=4S(\eps=1)+\cO (a^8)$ and does not allow us to change 
the sign of the $a^6$ term. The same holds for the Manton action~\cite{man32} 
which by definition agrees to $\cO (a^8)$ with the Wilson action.

In the past, more complicated improved actions were 
considered~\cite{sym8,wei25}, for which we will present the result similar to 
eq.(\ref{eq:epsac2}), as it allows us to predict whether or not they give rise 
to stable solutions~\cite{iwa7}. It also allows comparison with earlier results 
by L\"uscher and Weisz~\cite{wei25} obtained from a perturbative analysis. 
In the following, the coefficients in front of the $c_i$ are to match with the
definitions of ref.~\cite{wei25}. The averages $\langle\cdots\rangle$ are 
similar to the clover average above, but include now also averaging over 
all orientations of the loops. After some algebra one finds:
\bea
&&\hskip-5mm S(\{c_i\})\equiv\sum_x\Tr\{c_0\left\langle\sn\plaqa\,\right\rangle 
+2c_1\left\langle\sn\twooneplaq\,\right\rangle+4c_2\left\langle\sn\hookplaq
\,\right\rangle+\frac{4}{3}c_3\left\langle\sn\cornplaq \,\right\rangle\}=
\nonumber\\&&\hskip-5mm-\frac{a^4}{2}(c_0+8c_1+16c_2+8c_3)\sum_{x,\mu,\nu}
\Tr(F_{\mu\nu}^2(x))+a^6(c_2+\frac{c_3}{3})\sum_{x,\mu,\nu,\lambda}
\Tr(\cD_{\mu}F_{\mu\lambda}(x)\cD_{\nu}F_{\nu\lambda}(x))\nonumber\\
&&\hskip-5mm +\frac{a^6}{12}(c_0+20c_1+4c_2-4c_3)\sum_{x,\mu,\nu}\Tr(\cD_{\mu}
F_{\mu\nu}(x))^2+a^6\frac{c_3}{3}\sum_{x,\mu,\nu,\lambda}
\Tr((\cD_\mu F_{\nu\lambda})^2)+\cO(a^8)\ .
\label{eq:weisac}
\eea
One can therefore achieve tree-level improvement by
choosing~\cite{wei25} $c_0+8c_1+16c_2+8c_3=1$, $c_0+20c_1+4c_2-4c_3=0$ 
and $c_2=c_3=0$. Note that the condition $c_2+c_3/3=0$ only applies off-shell,
since on-shell $\sum_{x,\mu,\nu,\lambda}\Tr(\cD_{\mu}F_{\mu\lambda}(x)
\cD_{\nu}F_{\nu\lambda}(x))=0$.

Iwasaki and Yoshi\'{e}~\cite{iwa7} considered cooling for the Symanzik 
improved action, that is $c_0=\fthird,\ c_1=-\twth$ and $c_{2,3}=0$,
for which the $a^6$ term vanishes. The $a^8$ term 
will have to be computed to settle stability. From equation (\ref{eq:epsac2}) 
one sees that the $a^6$ term has a definite sign. This is no 
longer the case for the $a^8$ term. The same holds for the Symanzik 
improved action:
\bea
S_{\rm Symanzik}&=&\sum_{x,\mu,\nu}\Tr[-\frac{a^4}{2}F_{\mu\nu}^2(x)+
\frac{a^8}{24}\{F_{\mu\nu}^4(x)+\frac{1}{3}\cD_{\mu}^2F_{\mu\nu}(x)
\cD_{\nu}^2 F_{\mu\nu}(x)\nonumber\\ &-&\frac{1}{4}(\cD_{\mu}\cD_{\nu} 
F_{\mu\nu}(x))^2+\frac{4}{15}(\cD_{\mu}^2 F_{\mu\nu})^2\}]
+\cO(a^{10})
\quad.\label{eq:symac}
\eea
To decide in these cases if the lattice admits a stable solution (i.e. its
action increases with decreasing $\rho$), one can compute the lattice action
using explicitly the topological charge-one instanton solution with scale 
$\rho$. Eqs.(\ref{eq:epsac2},\ref{eq:weisac}) and (\ref{eq:symac}) are
only valid as long as $a\ll\rho$ because for $\rho\sim a$ the expansion 
in powers of $a$ no longer converges. For $\rho\ll L$ to a good approximation 
we can substitute the infinite volume continuum instanton solution:
\be
A_{\mu}(x)=-i\frac{\eta^a_{\mu\nu}x_{\nu}\sigma_a}{(x^2+\rho^2)}\quad,
\label{eq:inst}
\ee
with $\eta^a_{\mu\nu}$ the self-dual 't Hooft tensor~\cite{tho27}. When 
$\rho\sim L$ the solution will of course be modified by the boundary effect.
Substituting eq.(\ref{eq:inst}) we find
\bea
S(\eps)&=&8\pi^2\{1-\frac{\eps}{5}(a/\rho)^2 
-\frac{15\eps-12}{210}(a/\rho)^4+\cO(a/\rho)^6\} 
\quad,\nonumber\\
S_{\rm Symanzik} &=& 8\pi^2\{1-\frac{17}{210}(a/\rho)^4 
+\cO(a/\rho)^6\}\quad,\label{eq:epssymac}
\eea
we thus confirm the observation of Iwasaki and Yoshi\'{e}~\cite{iwa7} that the
Symanzik tree-level improved action has no stable instanton solutions. 
Since $S(\eps=0)=8\pi^2\{1+\frac{2}{35}(a/\rho)^4+\cO  
(a/\rho)^6\}$ we predict even at $\eps=0$ 
the lattice to have stable solutions, which we have verified for the case with 
twisted boundary conditions in the time direction (see below).

Iwasaki and Yoshi\'{e}~\cite{iwa7} also considered cooling for 
Wilson's choice~\cite{wil26} (W) of $c_0=4.376$, $c_1=-0.252$, $c_2=0$ 
and $c_3=-0.17$ and for (R) $c_0=9$, $c_1=-1$ and $c_2=c_3=0$. 
To $\cO(a^8)$ these actions effectively correspond respectively to 
$\eps=-2.704$ and $\eps=-11$, which for the case (W) we computed by 
substituting the continuum instanton solution. Indeed, they see stability 
up to 250 sweeps in both cases. 
\vskip1.5cm

\section{Non-leading lattice artefact corrections}

In presenting eq.(\ref{eq:epssymac}) we have replaced the sum over the lattice 
points by an integral and ignored the fact that on the lattice the equations of
motion are modified. Both effects turn out to be small, the first exponential 
in $\rho/a$, the other gives a correction to the expression for 
$S(\eps)$ in eq.(\ref{eq:epssymac}) proportional to 
$\eps^2(a/\rho)^4$ (whereas the correction to $S_{\rm Symanzik}$
is proportional to $(a/\rho)^8$, which also holds for 
$S(\eps=0)$).

We wish to compute $\sum_x f(x)=\sum_n f(na)$, for which we can use its 
Fourier decomposition
\be
a^4\sum_x f(x)=a^4\sum_{n\in \zahlen^4}\sum_k e^{ik\cdot na}\tilde{f}(k)=a^4 
N_s^3 N_t \sum_{p\in \zahlen^4}\tilde{f}(\frac{2\pi p}{a}) 
=\sum_{p\in \zahlen^4}\int e^{-2\pi p\cdot x/a} f(x) d^4x\quad.
\label{eq:sumint1}
\ee
The terms with $p\not=0$ give the error one makes, when replacing the lattice 
sum by an integral. For $a\ll\rho\ll L$ and $f(x)=-\frac{1}{2}\Tr
(F_{\mu\nu}^2(x+x_0))$ one finds explicitly (using eq.(\ref{eq:inst}))
\bea
-\frac{a^4}{2}\sum_{x,\mu,\nu}\Tr(F_{\mu\nu}^2(x+x_0)) & = & 8\pi^2[1+
\sum_{p\in \zahlen^4\setminus\{0\}}2\pi^2p^2(\rho/a)^2 cos(2\pi p\cdot x_0/a)
K_2(2\pi |p|\rho/a)] \nonumber\\
&=&8\pi^2[1-8\pi^2(\rho/a)^{3/2}e^{-2\pi\rho/a}(1+\cO(a/\rho))]\quad,
\label{eq:sumint2}
\eea
(with $K_2$ the modified Bessel function~\cite{abr28}). Here we have taken
$x_0$ to coincide with a point on the \underline{dual} lattice,
$2x_0^{\mu}=a$ for all $\mu$, as this minimizes the action.

To estimate the shift in the equations of motion due to the lattice artefacts,
we again consider $\rho\gg a$, such that the action can, in a good 
approximation, be given by
\be
\tilde{S}(\eps)=\sum_{\mu,\nu}\int d^4x \{-\frac{1}{2}\Tr
(F_{\mu\nu}^2(x))+\frac{\eps a^2}{12}\Tr((\cD_{\mu}F_{\mu\nu}
(x))^2)\}+\cO (a^4)\quad,\label{eq:twidac}
\ee
which implies the equations of motion:
\be
\sum_{\nu}\cD_{\nu}F_{\nu\mu}=\eps a^2 H_{\mu}\equiv -\eps 
a^2\sum_{\nu}(\frac{1}{12}\cD_{\nu}^3F_{\nu\mu} + \frac{1}{6}[F_{\mu\nu},
\cD_{\mu}F_{\mu\nu}] +\frac{1}{12}\cD^2_{\mu}\cD_{\nu}F_{\nu\mu})\quad.
\label{eq:el}
\ee
As eq.(\ref{eq:twidac}) breaks the scale invariance, there will in general not
be solutions close to eq.(\ref{eq:inst}). Variation with respect to $\rho$ 
no longer leaves the action invariant. Still, since this variation corresponds 
to a near zero-mode, it makes sense to expect quasi-stability under cooling. 
The action changes only slowly in the direction of this near zero-mode but
is predominantly lowered in those directions that 
leave the curvature square integrable and are spanned by the non-zero modes 
of the quadratic fluctuation operator for the action, which in the background 
gauge corresponds to 
\be
{\cal M}_{\lambda\sigma}=\delta_{\lambda\sigma}\cD^2_{\mu}+2{\rm ad}
F_{\lambda\sigma}\quad.\label{eq:quafluc}
\ee
If $\cP$ is the projection operator on the normalizable non-zero modes of
${\cal M}$ one has (at $\rho=1$)
\bea
A_{\mu}&=& A_{\mu}^{(0)}+\eps a^2\cP{\cal M}^{-1}_{\mu\nu}
\cP H_{\nu}\equiv  A_{\mu}^{(0)}+\eps a^2 A_{\mu}^{(1)}\quad,\nonumber\\
H_{\mu}(x) & = & \frac{16}{(1+x^2)^5}\psi_{\ \mu}^{(\thalf,\fhalf)}(x)
+\frac{8}{(1+x^2)^5}\psi_{\ \mu}^{(\half,\half)}(x)\quad,\label{eq:shift}
\eea
where $H_{\mu}(x)$ is evaluated by substituting for $A_{\mu}^{(0)}(x)$ 
the continuum solution $A_{\mu}(x)$ given in eq.(\ref{eq:inst}). 
For convenience we introduced the quantities $\psi_{\ \mu}^{(l,j)}(x)$:
\bea
\psi_{\ \mu}^{(\half,\half)}(x)&=&i\sum_{\nu,a}\eta^a_{\mu\nu}
x_{\nu}\sigma_a\quad, \quad
\psi_{\ \mu}^{(\thalf,\thalf)}(x)=i\sum_{\nu,a}\eta^a_{\mu\nu}
x_{\nu}\sigma_a(x^2-6x_{\mu}^2)\quad,\nonumber\\
\psi_{\ \mu}^{(\thalf,\fhalf)}(x)&=&i\sum_{\nu,a}\eta^a_{\mu\nu}
x_{\nu}\sigma_a(3x^2-3x_{\mu}^2-5x_{\nu}^2)\quad,
\label{eq:psifields}
\eea
which are eigenfunctions of the angular momentum operators $\vec{L}_1^2$ and
$\vec{J}^2$ (as defined in~\cite{tho27}  $L^a_1=-\frac{i}{2}\eta^a_{\mu\nu}
x_{\mu}\partial_{\nu}$, $J^a=L^a_1+ad(\frac{\sigma_a}{2})$). 
To compute ${\cal M}^{-1}_{\mu\nu}\cP H_{\nu}$ one can use for
${\cal M}^{-1}_{\mu\nu}$ the explicit expression~\cite{bro33} 
${\cal M}^{-1}_{\mu\nu}\equiv\bar{\eta}^a_{\mu\lambda}\cD_{\lambda}
(\cD_{\alpha}^{-2})\bar{\eta}^a_{\nu\sigma}\cD_{\sigma}$
(in the gauge $\cD_{\mu}(\cP H_{\mu})=0$). The result, which can 
be verified by applying ${\cal M}_{\mu\nu}$, is found to be 
\bea
{\cal M}^{-1}_{\mu\nu}\cP H_{\nu} 
&=&(\frac{\log(1+x^2)}{5(1+x^2)^2}-\frac{1}{3(1+x^2)^3}+\frac{1}{10(1+x^2)})
\psi^{(\half,\half)}_{\ \mu}(x) \nonumber\\
&+&(\frac{2\log(1+x^2)}{5x^8(1+x^2)^2}-\frac{6+3x^2-x^4}{15x^6(1+x^2)^3})
\psi^{(\thalf,\thalf)}_{\ \mu}(x) \nonumber\\
&+&(\frac{2(3+5x^2)\log(1+x^2)}{5x^8(1+x^2)^2}
-\frac{6+13x^2+4x^4}{5x^6(1+x^2)^3})\psi^{(\thalf,\fhalf)}_{\ \mu}(x)
\quad.\label{eq:nonnormA1}
\eea
The algebraic manipulation program Mathematica~\cite{wol34} was useful in
 obtaining and checking these results. Despite its appearance, this result is 
regular at $x\rightarrow0$. However, it contains a non-normalizable 
deformation (since $\psi^{(\half,\half)}_{\ \mu}(x)/(1+x^2)=-A_{\mu}^{(0)}(x)$),
which would make the action diverge and should be removed by projecting on 
normalizable deformations:
\be
A_{\mu}^{(1)}=\cP{\cal M}^{-1}_{\mu\nu}\cP H_{\nu}= 
{\cal M}^{-1}_{\mu\nu}\cP H_{\nu} -\frac{(23+17x^2)}{60(1+x^2)^2}
\psi^{(\half,\half)}_{\ \mu}(x),\quad
{\cal M}_{\mu\nu}A^{(1)}_{\nu}=H_{\mu}-\frac{8
\psi^{(\half,\half)}_{\ \mu}(x)}{5(1+x^2)^3}.\label{eq:normA1}
\ee
One easily verifies that $A_{\mu}^{(1)}$ and ${\cal M}_{\mu\nu}A_{\nu}^{(1)}$
are both square integrable and orthogonal to the zero-mode
$\partial A_{\mu}^{(0)}/\partial\rho|_{\rho=1}=2\psi^{(\half,\half)}_{\ \mu}(x)
/(1+x^2)^2$.

We can now substitute $A_{\mu}=A_{\mu}^{(0)}+\eps a^2 A_{\mu}^{(1)}$
in eq.(\ref{eq:epsac2}) (for later convenience evaluated for a different
$\eps$) to obtain the shift in the action
\be
\tilde{S}(\tilde{\eps})=S(\tilde{\eps})-2\eps
\tilde{\eps}a^4\int d^4x\Tr(A^{(1)}_{\mu}(x)H_{\mu}(x))
+\eps^2 a^4 \int d^4x \Tr(A^{(1)}_{\mu}(x){\cal M}_{\mu\nu}
A^{(1)}_{\nu}(x))+\cO(a^6)\quad,\label{eq:twidacpert}
\ee
where for the term linear in $A^{(1)}_{\mu}$ we used the equations of motion 
for $A^{(0)}_{\mu}$. We also reintroduced the $\rho$ dependence using
trivial dimensional arguments. Evaluating the integrals gives
\be
\tilde{S}(\tilde{\eps})=8\pi^2\{1-\frac{\tilde{\eps}}{5}
(\frac{a}{\rho})^2 -[\frac{15\tilde{\eps}-12}{210} +\frac{284\eps
\tilde{\eps}}{2625} -\frac{179\eps^2}{5250}](\frac{a}{\rho})^4 
+\cO(\frac{a}{\rho})^6\}\quad.\label{eq:twidacexp}
\ee
At $\eps=\tilde{\eps}=-1$ the shift due to the modified 
equations of motion in the $(a/\rho)^4$ term is 58\%.

Strictly speaking our expression for the $\rho$-dependence of the lattice
action is only valid for $\rho/a\stackrel{>}{\sim}2$ and $\rho\ll L$,
since we are using the continuum infinite volume solution as the zero-order 
approximation. But even for $\rho\sim L/2$ it is not unreasonable to expect
the order of magnitude of the corrections to be given by eqs.(\ref{eq:sumint2})
and (\ref{eq:twidacexp}).

\section{Numerical results and discussion}

This section discusses the numerical results obtained as described in section 
3, mainly to illustrate the viability of our ideas. A more detailed and 
careful analysis will be left for a future publication. So far we have worked 
mainly on lattices of size $N_s^3\times N_t$, with $N_s=7$ or $8$ and with 
$N_t=3N_s$ to $4N_s$. At $\eps=-1$ (see eq.(\ref{eq:epsac1})) we settle to an 
action near $8\pi^2$, and we have seen stability for up to 6000 sweeps. The 
same is true for $\eps=0$ in the presence of a twist (but not without twist, 
where our configuration ultimately decays to the vacuum at $\eps=0$, but note 
that in that case there are no regular instanton solutions). Apart from the 
total action we compute separately the sum over the $n\times m$ plaquettes, 
denoted by $S_{n\times m}$ and averaging over the two orientations if 
$n\neq m$. $S_{n\times m}$ is normalized by dividing by $8\pi^2n^2m^2$, such
that for an infinite lattice and $\rho/a\rightarrow\infty$,
$S_{n\times m}\rightarrow 1$. When we perform cooling with the action of
eq.(\ref{eq:epsac1}) we should take $A_\mu=A_\mu^{(0)}+\eps a^2 
A_\mu^{(1)}$ in calculating $S_{n\times m}$. Eq.(\ref{eq:clover}) for 
$S_{1\times 1}$ easily leads to the general result for $S_{n\times  m}$  by 
inserting for each index $\mu(\nu)$ a factor $n(m)$. Together with 
eq.(\ref{eq:twidacpert}) one deduces, that to $\cO(a^6)$
\bea
S_{n\times m}&=&1-\frac{(n^2+m^2)}{2}\alpha a^2-\left(m^2n^2\beta_1-
\frac{m^4+n^4}{2}\beta_2+\frac{n^2+m^2}{2}\eps\gamma-\eps^2\delta\right)a^4
\ ,\label{eq:snm}
\eea
up to the discretization error implied by eq.(\ref{eq:sumint2}),
which for the lattices we are considering can be estimated to be not bigger 
than $10^{-6}$.
This formula holds for sufficiently smooth configurations, i.e. 
$\alpha a^2\ll 1$, even if the configuration has non-vanishing action over 
the entire spatial volume. It is these configurations that are of interest 
to us and which deviate considerably from localized instantons
(eq.(\ref{eq:inst})) for which $\rho \ll L$. From eqs. (\ref{eq:epssymac},
\ref{eq:twidacexp}) we easily deduce for those localized instantons the 
results:
\be
\alpha a^2=\frac{1}{5}(\frac{a}{\rho})^2,\quad\beta_1 a^4=\frac{29}{630} 
(\frac{a}{\rho})^4,\quad\beta_2 a^4=\frac{2}{63}(\frac{a}{\rho})^4,\quad
\gamma a^4=\frac{284}{2625}(\frac{a}{\rho})^4,\quad\delta a^4=\frac{179}{5250}
(\frac{a}{\rho})^4.
\label{eq:ana}
\ee From the numerical results we have obtained $S_{1\times 1}$, 
$S_{1\times 2}$, $S_{2\times 2}$ and $S_{1\times 3}$ on two lattices of size
respectively $7^3\times 21$ and $8^3\times 24$, for $\eps=0$ and $\eps=-1$ 
with a twist $n_{0i} = (1,1,1)$
(see table I). From these we extract the coefficients in 
eq.(\ref{eq:snm}), whose values are summarized in table II (the 
error due to neglecting the $O(a^6)$ term is of the order of 
$(n^6+m^6)(\alpha a^2)^3$).
\vskip3mm
\hbox to \hsize{\hfil\vbox{\offinterlineskip
\halign{&\vrule#&\ $#\mystrut$\hfil\ \cr
\noalign{\hrule}
&N_s\times N_t&&\phm\eps&&\quad S_{1\times 1}&&\quad S_{1\times 2}
&&\quad S_{2\times 2}&&\quad S_{1\times 3}&\cr
height 4pt&\omit&&\omit&&\omit&&\omit&&\omit&&\omit&\cr
\noalign{\hrule}
height 4pt&\omit&&\omit&&\omit&&\omit&&\omit&&\omit&\cr
&7^3\times 21&&   -1&&0.982591&&0.957050&&0.928823&&0.918105&\cr
&7^3\times 21&&\phm0&&0.982287&&0.956437&&0.927908&&0.917109&\cr
&8^3\times 24&&   -1&&0.986720&&0.967122&&0.945887&&0.936619&\cr
&8^3\times 24&&\phm0&&0.986529&&0.966736&&0.945310&&0.935976&\cr
\noalign{\hrule}}}\hfil}
\vskip3mm
{\narrower{\noindent
Table I: Numerical results obtained by cooling with $S(\eps)$
and twist $n_{0i}=(1,1,1)$.}\par}
\vskip5mm
\hbox to\hsize{\hfil\vbox{\offinterlineskip\halign
{&\vrule#&\ $#\mystrut$\hfil\ \cr \noalign{\hrule}
&N_s\times N_t&&\quad \alpha a^2&&\beta_1/\alpha^2&&\beta_2/\alpha^2
&&\gamma/\alpha^2&&\delta/\alpha^2&\cr
height 4pt&\omit&&\omit&&\omit&&\omit&&\omit&&\omit&\cr
\noalign{\hrule}
height 4pt&\omit&&\omit&&\omit&&\omit&&\omit&&\omit&\cr
&7^3\times 21&&0.01761&&0.96&&0.63&&0.66&&0.32&\cr
&8^3\times 24&&0.01340&&1.01&&0.64&&0.71&&0.35&\cr
&\rho<<L&&0.2(a/\rho)^2&&1.15&&0.79&&2.70&&0.85&\cr
\noalign{\hrule}}}\hfil}
\vskip3mm
{\narrower\narrower{\noindent
Table II: Coefficients appearing in eq.(\ref{eq:snm}) extracted from the 
numerical \\ results in table I, using $S_{1\times1}$, $S_{2\times2}$ and 
$S_{1\times2}$ (the latter at $\eps=-1$ only).}\par}

It is interesting to analyse the untwisted case in more detail to illustrate 
the difficulty in having self-dual solutions at finite $T$. In figure 1 we 
plot (a) the total electric and magnetic energies $\cE_{E,B}(t)$, (b)
the Polyakov-line $P_1(t)$ through two particular points $\vec{x}$ and 
similarly for $P_{2,3}(t)$ in (c,d). We see two features that are intimately 
related. First, where $\cE_B(t)\rightarrow 0$, the electric energy 
$\cE_E(t)\rightarrow$ const. Second for the same $t$ values where this 
occurs $C_i(t)$ ($P_i(t)=\cos(C_i(t)/2)$) is linear in $t$ and $\vec{x}$ 
independent. These are precisely the equations of motion when restricting 
to the vacuum valley. Classically motion on this valley, which itself has 
the geometry of a three-torus, is free. On the lattice this motion is 
described by the action
\be
\sum_t 4N_s^3(1-\cos(\frac{C_k(t+1)-C_k(t)}{2N_s}))\quad.\label{eq:tail}
\ee
One easily checks that the values of $C_k(t+1)-C_k(t)$ obtained from 
figs.1(b,c,d) quite accurately reproduce through eq.(\ref{eq:tail}) the 
value for $\cE_E(t)$. Clearly the electric tail destroys the self-duality. 
Suppose that at $T\rightarrow \infty$ the solution describes tunnelling 
from $C_i^-$ to $C_i^+$ and $C_i^+\neq C_i^-$, then at finite $T$ the periodic 
boundary conditions force $C_i(t)$ to linearly interpolate between $C_i^+$ 
and $C_i^-$ over a time $T-T_0$, if $T_0$ is the time interval over which 
$\cE_B(t)\neq 0$. Thus the action, even in the continuum, would be bigger 
than $8\pi^2$ by a number proportional to $1/(T-T_0)$ except when there are 
solutions with $C_i^-=C_i^+$ for $T\rightarrow \infty$. These can certainly 
not be excluded, in particular as $C_i^+=C_i^-=\pi$ is compatible with a twist 
$n_{0i}=1$, but if these are very localized instantons, the lattice artefacts 
might make its action so big, that the lattice will prefer the least localized 
solutions with $C_i^+\neq C_i^-$. If $T$ is not big enough the lattice will 
find a compromise between these two cases. There are indications that the 
largest instanton prefers  $C_i^+\neq C_i^-$ and from the numerical results 
with twist $n_{0i}=(1,1,1)$ presented in fig.2, the preferred 
values seem to be such that two of the $C_i$ go from $0$ to $2\pi$ and one 
goes from $2\pi/3$ to $4\pi/3$. We compare (after appropriate scaling with 
$N_s$) for $N_s=7$ and 8, using over-improved cooling at $\eps=-1$, in fig.2a 
the electric and magnetic energy profiles, and in figs.2b-d the values of 
$C_i\equiv2{\rm acos}(P_i)$, at the spatial lattice point with maximal energy
(to be precise, with maximal $E_1^2$). From this we deduce $\cE_E=\cE_B$ to a 
high accuracy, consistent with self-duality, and the excellent scaling with 
$N_s$. The results in figure 2 are obtained after roughly 6600 cooling sweeps,
which is necessary since the dependence of the lattice action on $C_i^{\pm}$
is rather weak (at $\eps=0$ too weak to observe)
and the configuration only slowly reaches the minimum of the 
lattice action. We have verified that the approach to this minimum is 
exponential, as is illustrated in fig.3, where we plot the total action
and the maximum of $\cE_B(t)$ (i.e. $\cE_B(0)$) as a function of the number of 
cooling sweeps. We see indeed that the maximal energy along the tunnelling
path decreases under cooling, which is mainly due to the increasing size, as
otherwise the action should depend more strongly on the number of cooling
sweeps. (For the Wilson action one sees a dramatic increase of $\cE_B(0)$
under cooling, until the action suddenly drops to zero.)
With boundary conditions that 
fix the link variables at $t=0$ and $t=T$ to the vacuum configurations,
the approach to the minimum action is much faster. 

Elsewhere we will publish a more detailed analysis of the scaling properties,
as well as testing our conjecture to be able to find a sphaleron. Also 
numerical results with fixed boundary conditions, that allow us to investigate 
if solutions exist for arbitrary $C_i^\pm$ will be presented elsewhere. 
This paper mainly served the purpose to describe the formalism, and 
demonstrate the large amount of control obtained in this way in studying 
instanton solutions on a torus.

It would be interesting to repeat this analysis for the 2-dimensional O(3) 
model, for which the instantons on a torus are exactly known~\cite{ric35}, 
in the light of the ``perfect'' lattice action recently considered by 
Hasenfratz and Niedermayer~\cite{has36}. But as we have shown, appropriate 
deviations from a ``perfect'' action can be quite helpful.

Finally, over-improvement might be an efficient tool to measure the topological 
susceptibility, as the action to generate a statistical ensemble need not 
be the same as the one used to measure the topological charge.

\section{Acknowledgements}

We have benefited from discussions with Wolfgang Bock, Maarten Golterman,
Peter Hasenfratz, Jim Hetrick, Arjan Hulsebos, Ferenc Niedermayer, Erhard
Seiler and Peter Weisz. In particular we thank Peter Braam and Jan Smit for 
discussions on respectively the moduli spaces and improved actions. One
of us (P.v.B.) wishes to take this opportunity to thank Erhard Seiler and
Peter Weisz for hospitality at the Max Planck Institute in Munich, Dan
Freedman and Ken Johnson for hospitality at MIT and Tony Gonz\'alez-Arroyo
for hospitality at UAM in Madrid.

This work was supported in part by grant number AEN 90-0272, financed by CICYT,
and by grants from ``Stichting voor Fundamenteel Onderzoek der Materie (FOM)'' 
and ``Stichting Nationale Computer Faciliteiten (NCF)'' for use of the CRAY 
Y-MP at SARA. M.G.P. was supported by a Human Capital and Mobility EC 
fellowship.

\eject
\begin{center}
\Large{\bf Figure captions}
\end{center}
\vskip1cm
{\narrower\narrower{\noindent
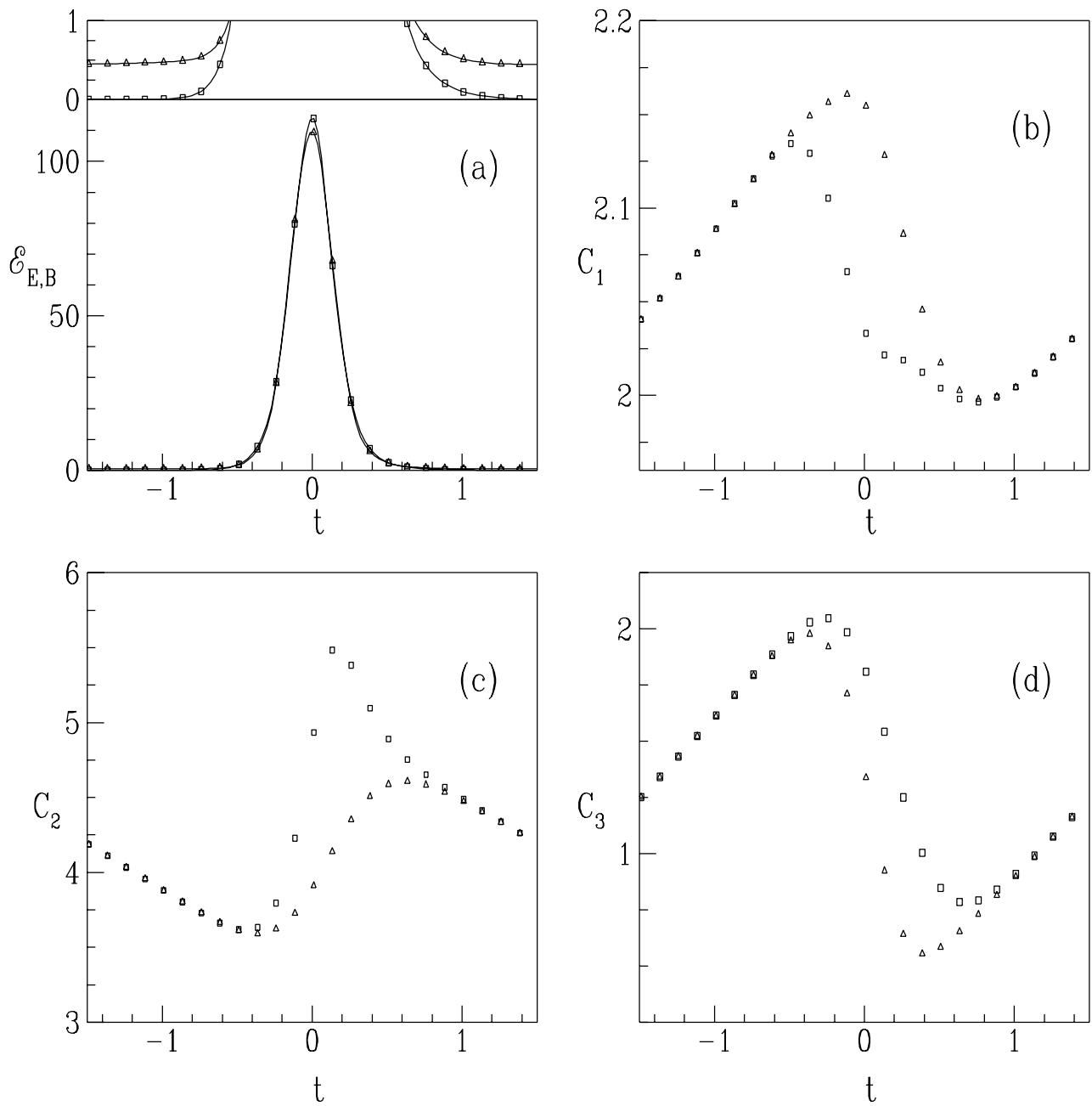
Figure 1: Numerical results (after scaling appropriately with $N_s$)
for the case of an $8^3\times 24$ lattice without twist, obtained from
over-improved cooling at $\eps=-1$. In (a) the electric ($\cE_E(t)$ triangles) 
and magnetic ($\cE_B(t)$ squares) energies are plotted. In the upper part 
of this figure the tails are plotted at an enlarged scale. In (b-d) are 
plotted $C_i(t)\equiv2{\rm acos}(P_i(t))$ through two distinct spatial 
points on the lattice.}\par}
\vskip1cm
{\narrower\narrower{\noindent
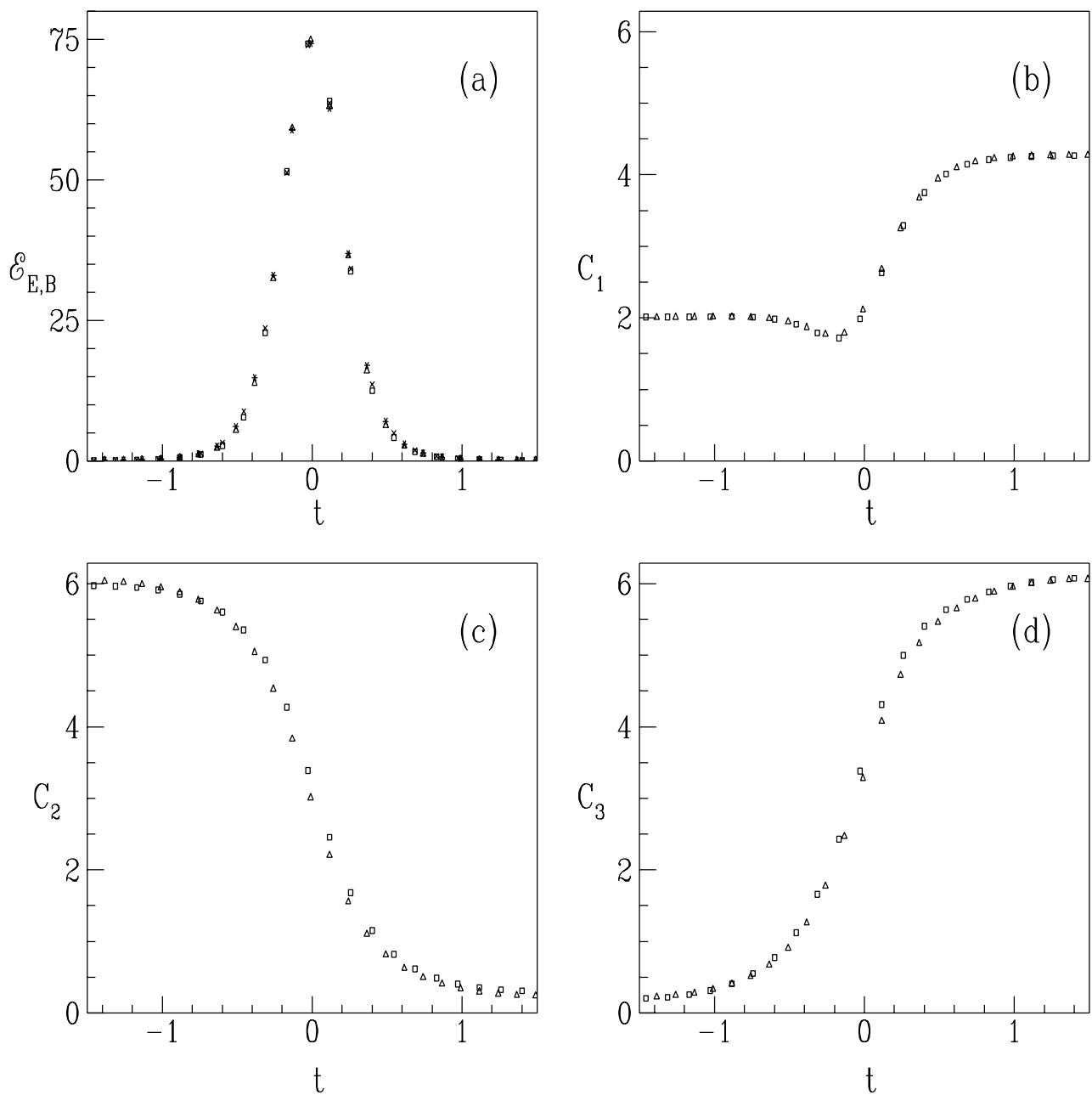
Figure 2: Numerical results (after scaling appropriately with $N_s$)
for the cases of a $7^3\times 21$ (squares) and an $8^3\times 24$ (triangles)
lattice with twist $n_{0i}=(1,1,1)$, obtained from over-improved cooling 
at $\eps=-1$. Figure (a) contains four data sets. Two for $\cE_E(t)$ with the 
above mentioned symbols and two (crosses for $N_s=7$ and stars for $N_s=8$)
for $\cE_B(t)$. Figures (b-d) exhibit $C_i(t)$, through the spatial lattice 
point with maximal $E_1^2$ at $t=0$.}\par}
\vskip1cm
{\narrower\narrower{\noindent
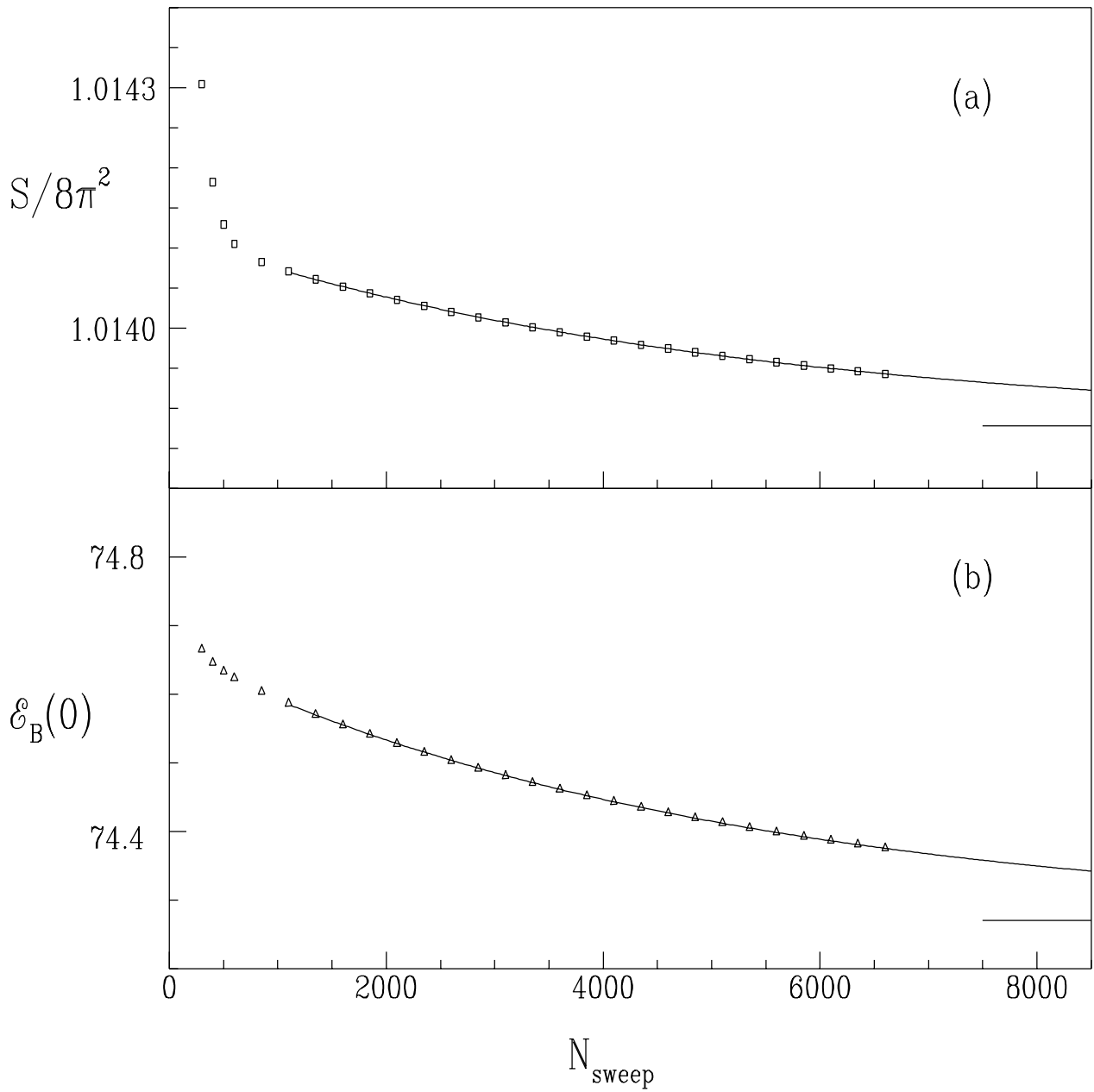
Figure 3: The history of the action $S(\eps=-1)$ and the maximal magnetic 
energy $\cE_B(t=0)$ as a function of the number of cooling sweeps for an 
$8^3\times 24$ lattice with twist $n_{0i}=(1,1,1)$, together with their 
exponential fits. The short lines on the right indicate the asymptotic values 
following from these fits.}\par}
\end{document}